\documentclass{iau}

\newcommand{\projecttitle}{Carving out the low surface brightness universe with NoiseChisel}
\newcommand{\zenododoi}{https://doi.org/10.5281/zenodo.6529419}
\newcommand{\projectversion}{751467d}
\newcommand{\projectgitrepo}{https://gitlab.com/makhlaghi/iau-symposium-355}
\newcommand{\projectcopyrightowner}{Mohammad Akhlaghi <mohammad@akhlaghi.org>}
\newcommand{\maneagedate}{4 May 2022}
\newcommand{\maneageversion}{f51b5e2}
\newcommand{\firstsubmitdate}{25 Sep 2019}
\newcommand{\firstsubmitcommit}{8505cfd}

\newcommand{\uvudfrdoneid}{6331}
\newcommand{\uvudfrdtwoid}{6338}
\newcommand{\uvudfrconeid}{6313}
\newcommand{\uvudfrctwoid}{6460}
\newcommand{\uvudfrsoneid}{6341}
\newcommand{\uvudfrstwoid}{6431}

\newcommand{\gnuastroversion}{0.10}
\newcommand{\sextractorversion}{2.25.0}
\newcommand{\demosfdefaultsn}{0.59}
\newcommand{\demosfdefaultmagperarc}{25.04}
\newcommand{\demosfoptimizedsn}{0.25}
\newcommand{\demosfoptimizedmagperarc}{25.97}
\newcommand{\sdsszpdiff}{2.3}
\newcommand{\sdsszpasinhr}{24.80}
\newcommand{\sdsszpnanomaggie}{22.50}
\newcommand{\ncsnquant}{0.95}
\newcommand{\nctilesize}{75,75}
\newcommand{\ncminskyfrac}{0.9}
\newcommand{\ncinterpnumngb}{6}
\newcommand{\ncdetgrowquant}{0.65}
\newcommand{\ncmeanmedqdiff}{0.001}
\newcommand{\ncdetgrowmaxholesize}{10000}
\newcommand{\sedetectthresh}{0.5}
\newcommand{\seanalysisthresh}{0.5}

\newcommand{\machinearchitecture}{x86\_64}
\newcommand{\machinebyteorder}{Little Endian}

\ifdefined\highlightnew

\else

\fi

\ifdefined\highlightnotes
\newcommand{\tonote}[1]{\textcolor{red!60!black}{[#1]}}
\else
\newcommand{\tonote}[1]{{}}
\fi

\usepackage{graphicx}
\usepackage{xcolor}
\usepackage[hyphens]{url}
\usepackage{hyperref}
\hypersetup{
     colorlinks  = true,
     linkcolor   = blue,
     anchorcolor = red,
     citecolor   = blue,
     filecolor   = blue,
     menucolor   = blue,
     runcolor    = cyan,
     urlcolor    = blue,
     linktocpage = true,
     pdftitle={\projecttitle},
     pdfauthor={\projectcopyrightowner},
     pdfsubject={\projectgitrepo{} (commit \projectversion)},
     pdfkeywords={NoiseChisel, Gnuastro, Detection, Low surface brightness, Reproducible research, Maneage}
}

\newcommand{\segment}{\textsf{Segment}}
\newcommand{\snsign}{{\small S}/{\small N}}
\newcommand{\sextractor}{\textsf{SE\-xtractor}}
\newcommand{\noisechisel}{\textsf{Noise\-Chisel}}

\newcommand{\snsignsmaller}{{\footnotesize S}/{\footnotesize N}}

\usepackage{fixltx2e}

\usepackage{xcolor}
\color{black} 
\definecolor{DarkBlue}{RGB}{0,0,90}

\usepackage{setspace, caption}
\usepackage[normalem]{ulem}

\usepackage{csquotes}

\usepackage{xcolor}

\usepackage[
    doi=false,
    url=false,
    dashed=false,
    eprint=false,
    maxbibnames=2,
    minbibnames=1,
    hyperref=true,
    maxcitenames=2,
    mincitenames=1,
    giveninits=true,
    style=authoryear,
    uniquelist=false,
    backend=biber,natbib]{biblatex}
\DeclareFieldFormat[article]{pages}{#1}
\DeclareFieldFormat{pages}{\mkfirstpage[{\mkpageprefix[bookpagination]}]{#1}}
\renewbibmacro{in:}{}
\AtEveryBibitem{\clearfield{month}}
\renewcommand*{\bibfont}{\footnotesize}
\DefineBibliographyStrings{english}{references = {References}}

\DeclareSourcemap{
  \maps[datatype=bibtex]{
    \map{
      \step[fieldsource=adsurl,fieldtarget=iswc]
      \step[fieldsource=gbkurl,fieldtarget=iswc]
    }
  }
}

\definecolor{mypurp}{cmyk}{0.75,1,0,0}
\newcommand{\doihref}[2]{\href{#1}{\color{magenta}{#2}}}
\newcommand{\adshref}[2]{\href{#1}{\color{mypurp}{#2}}}

\DeclareFieldFormat{doilink}{
  \iffieldundef{doi}{#1}{\doihref{http://dx.doi.org/\thefield{doi}}{#1}}}
\DeclareFieldFormat{adslink}{
    \iffieldundef{iswc}{#1}{\adshref{\thefield{iswc}}{#1}}}
\DeclareFieldFormat{arxivlink}{
  \iffieldundef{eprint}{#1}{\href{http://arxiv.org/abs/\thefield{eprint}}{#1}}}

\DeclareListFormat{doiforbook}{
  \iffieldundef{doi}{#1}{\doihref{http://dx.doi.org/\thefield{doi}}{#1}}}
\DeclareFieldFormat{googlebookslink}{
    \iffieldundef{iswc}{#1}{\adshref{\thefield{iswc}}{#1}}}

\DeclareBibliographyDriver{article}{%
  \usebibmacro{bibindex}%
  \usebibmacro{begentry}%
  \usebibmacro{author/translator+others}%
  \newunit%
  \ifdefined\makethesis\printtext{\usebibmacro{title}}\fi%
  \newunit%
  \printtext[doilink]{\usebibmacro{journal}}%
  \addcomma%
  \printtext[adslink]{\printfield{volume}}%
  \addcomma%
  \printtext[arxivlink]{\printfield{pages}}%
  \addperiod%
}

\DeclareBibliographyDriver{book}{%
  \usebibmacro{bibindex}%
  \usebibmacro{begentry}%
  \usebibmacro{author/translator+others}%
  \newunit%
  \printtext{\usebibmacro{title}}%
  \addperiod%
  \addspace%
  \printlist[doiforbook]{publisher}%
  \addcomma%
  \addspace%
  \printfield[googlebookslink]{edition}%
  \printtext{ ed.}%
  \addperiod%
}

\usepackage{tikz}
\usetikzlibrary{external}
\tikzsetexternalprefix{tex/tikz/}

\usepackage{pgfplots}
\pgfplotsset{compat=newest}
\usepgfplotslibrary{groupplots}
\pgfplotsset{
  axis line style={thick},
  tick style={semithick},
  tick label style = {font=\footnotesize},
  every axis label = {font=\footnotesize},
  legend style = {font=\footnotesize},
  label style = {font=\footnotesize}
  }

\pubyear{2019}
\volume{355}
\jname{The Realm of the Low-Surface-Brightness Universe}
\editors{D. Valls-Gabaud, I. Trujillo \& S. Okamoto, eds.}

\title[Carving out the low surface brightness universe]{\projecttitle}
\author[Mohammad Akhlaghi] {Mohammad Akhlaghi$^{1,2}$}
\affiliation{$^1$Instituto de Astrof\'isica de Canarias, C/ V\'ia L\'actea,
  38200 La Laguna, Tenerife, Spain. \\
  email: {\tt mohammad@akhlaghi.org} \\
  [\affilskip] $^2$Facultad de F\'isica, Universidad de La Laguna, Avda.
  Astrof\'isico Fco. S\'anchez s/n, 38200\\ La Laguna, Tenerife, Spain.}

\begin{document}

\maketitle
\begin{abstract}
  \noisechisel{} is a program to detect very low signal-to-noise ratio (\snsignsmaller) features with minimal assumptions on their morphology.
  It was introduced in 2015 and released within a collection of data analysis programs and libraries known as GNU Astronomy Utilities (Gnuastro).
  The 10th stable version of Gnuastro was released in August 2019 and {\noisechisel} has significantly improved: detecting even fainter signal, enabling better user control over its inner workings, and many bug fixes.
  The most important change until version 0.10 is that \noisechisel's segmentation features have been moved into a new program called \segment.
  Another major change is the final growth strategy of its true detections, for example \noisechisel{} is able to detect the outer wings of M51 down to \snsignsmaller{} of $\demosfoptimizedsn$, or $\demosfoptimizedmagperarc$ mag/arcsec$^2$ on a single-exposure SDSS image (r-band).
  \segment{} is also able to detect the localized HII regions as ``clumps'' much more successfully.
  For a detailed list of improvements after version 0.10, see the most recent manual.
  Finally, to orchestrate a controlled analysis, the concept of reproducibility is discussed: this paper itself is exactly reproducible (commit \texttt{\projectversion}\footnote{\url{https://gitlab.com/makhlaghi/iau-symposium-355}}).

\end{abstract}

\firstsection 
\section{Introduction}
Signal from the low surface brightness universe is buried deep in the dataset’s noise and thus requires accurate detection methods.
In \citet{gnuastro} (henceforth AI15) a new method was introduced to detect such very low signal-to-noise ratio (S/N) signal from the images in a non-parametric manner.
It allows accurate detection of the diffuse outer features of galaxies (that often have a different morphology from the centers).
The software implementation of this method (\noisechisel) is released as part of a larger collection of data analysis software known as GNU Astronomy Utilities\footnote{\url{https://www.gnu.org/s/gnuastro}} (Gnuastro).
It was the first professional astronomical software to be refereed by an independent panel (GNU Evaluation committee) and fully conforms with the GNU Coding Standards\footnote{\url{https://www.gnu.org/prep/standards}}.

\begin{figure}[t]
  \ifdefined\makepdf%
    \tikzsetnextfilename{uvudf}%
    \input{tex/src/uvudf.tex}%
  \else
    \includegraphics[width=\linewidth]{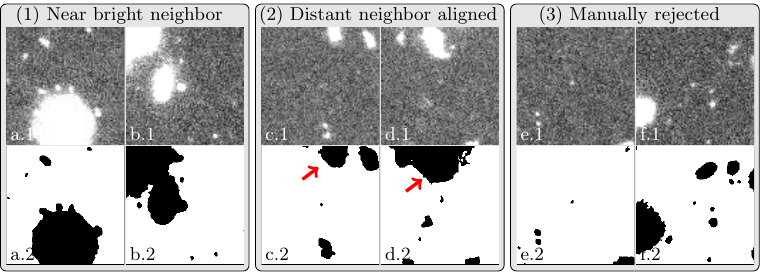}
  \fi

  \setlength\belowcaptionskip{1mm}
  \caption{\label{uvudf}
    Example targets (central object in top row) found in MUSE datacubes \citep{inami17} (I17) that did not correspond to anything in the \citet{rafelski15} (R15) catalog which is based on HST images.
    They had footprints in the combined \sextractor{} segmentation map of R15 (bottom row), so they are `detectable'.
    However, they were missed due to incorrect labeling: a segmentation/deblending problem.
    The top row is the HST/ACS F775W image from the XDF survey \citep{illingworth13}.
    From left to right, their IDs in I17 are \uvudfrconeid, \uvudfrctwoid, \uvudfrdoneid, \uvudfrdtwoid, \uvudfrsoneid, \uvudfrstwoid.
    In case (2), the central detection has the same label as the large detection that the red arrow is pointing to.
    Note that the R15 segmentation shown is a combination of six \sextractor{} runs with different deblending parameters.
    For more, see Section 7.3 of \citet{bacon17}.
  }
\end{figure}

Since its release, \noisechisel{} has been used in many studies.
For example \citet{bacon17} used it to identify objects that were missed by \citet{rafelski15} (henceforth R15), who used a combination of six \textsf{SExtractor} \citep{sextractor} runs with different configurations to avoid deblending problems, but still missed many sources with significant signal, see Figure \ref{uvudf}.
\citet{borlaff19,muller19,trujillo19} used it for accurate flat field and Sky subtraction to create deeper co-added images in galaxy fields for optimal detection of the low surface brightness features.
\citet{calvi19} used it to find Lyman-$\alpha$ emitters in spectra.
For future studies, \citet{laine18,hsieh19} have proposed using \noisechisel{} to maximize LSST's low surface brightness features (in extra galactic and solar system science outputs).

With ten stable releases until this publication, \noisechisel{} and Gnuastro have greatly evolved.
In Sections \ref{updatednc} and \ref{updatedseg} a review of the major changes and improvements in Gnuastro's detection features since its first release is discussed and Section \ref{m51demo} demonstrates these new features on an example image of M51.
Finally, recording and publishing the analysis (i.e., reproducibility) is particularly important for low surface brightness science, therefore AI15 was also built in a reproducible manner.
A short review of the evolved work in this frontier is given in Section \ref{repanalysis}.

\section{\noisechisel{} improvements}\label{updatednc}

Over the first ten stable releases of Gnuastro (version \gnuastroversion) \noisechisel{} has also greatly improved (as with many of Gnuastro's programs): some of the original features have been removed and many new features have been added.
In Subsections \ref{noisechiselremoved} and \ref{noisechiselnew} the removed and new features of \noisechisel{} are respectively discussed.
Note that \noisechisel{} will continue to evolve after this paper also, so the best reference to consult is always its own manual/book, in particular the ``\noisechisel{} changes after publication'' section.

The overall direction of Gnuastro's evolution (including \noisechisel) has been modularity and more user control.
In terms of modularity, the most notable change has been that \noisechisel{} doesn't do segmentation any more.
Segmentation of the signal into its substructure is fundamentally a separate kind of operation than detection (or separating the noise from the significant signal).

In terms of more user control \noisechisel{} has gradually moved in the direction of not having any number/configuration hard-coded into it.
This gives the user a very detailed control over how it operates at every step.
As a scientific software, it is critically important for the user to be able to tweak any step and understand the systematics of their final choice.
In the second and third tutorials of the Gnuastro manual/book the logic behind customizing them to fit different noise properties is fully described in real analysis scenarios.

\subsection{\noisechisel{} removed features}\label{noisechiselremoved}
\begin{itemize}
\item  Segmentation has been wholly moved to a new program; \segment. From Gnuastro 0.6, \noisechisel{} is only in charge on detection.
  Therefore its output is now just a binary/two-value image (0: Noise or Sky, 1: Signal).
  Identifying the sub-structure over the detected regions (for example true clumps) is a fundamentally different problem.
  Changes in the original segmentation steps are discussed in Section \ref{updatedseg}.
  This spin-off allows much greater modularity/creativity and is in the spirit of Gnuastro's modular design.

  For example in some scenarios there may be no Sky (the whole dataset is covered by targets) and detection would thus not be necessary.
  This happens when imaging parts of dense resolved objects for example globular clusters or the wings of M31.
  In other cases, segmentation isn't necessary, for example to estimate the Sky or flat-field level in individual exposures during data reduction like \citet{borlaff19}.

\item \texttt{--skysubtracted}: This option was used to account for the extra noise that is added if the Sky value has already been subtracted.
  However, \noisechisel{} estimates the Sky standard deviation based on the input data, not exposure maps or other theoretical considerations.
  Therefore the standard deviation of the undetected pixels also contains the errors due to any previous sky subtraction. This option is therefore no longer present in \noisechisel.

\item \texttt{--dilate}: During detection, the outer layers of true detections are eroded.
  Until version 0.5, after the true detections were found, they were dilated to recover the eroded pixels.
  However, the simple 8- or 4-connected dilation produced boxy/diamond shapes respectively.
  They were thus were not realistic and would miss diffuse flux.
  The final dig into the noise is now done by ``grow''ing the true detections.
  Similar to how true clumps were grown, see the description of \texttt{--detgrowquant} below.
\end{itemize}

\subsection{\noisechisel{} new features}\label{noisechiselnew}
The new features in \noisechisel{} are listed below. Note that their order is based on \noisechisel's functionality: which step is done before which.
\begin{itemize}
\item
  The quantile difference to identify tiles with no significant signal is measured between the \emph{mean} and median.
  Until Gnuastro 0.8, it was between the \emph{mode} and median.
  The quantile of the mean is more sensitive to skewness (the presence of signal), so it is preferable to the quantile of the mode.

\item
\texttt{--widekernel}: By default, \noisechisel{} convolves the input once and estimates the proper tiles for quantile estimations on the convolved image.
The same convolved image is later used for quantile estimation.
A larger kernel has the advantage that it increases the skewness (and thus difference between the mean and median, therefore helps in detecting the presence signal).
However, a larger kernel disfigures the shapes/morphology of the objects.

The new \texttt{--widekernel} option (and a corresponding \texttt{--wkhdu}
option to specify its HDU) were added in Gnuastro 0.5 to allow benefiting from the advantage and not being affected by the problem.
When it is given, the input will be convolved with both the sharp kernel (given through the \texttt{--kernel} option) and wide kernel (given through this option).
The mean and median are calculated on the dataset that is convolved with the wider kernel and good tiles are chosen.
Finally, quantiles are estimated on the good tiles based on the convolution with the sharper kernel.

\item
  Rejection of outlier tiles used in estimating the quantile threshold: when there are large galaxies or bright stars in the image, their gradient may be on a smaller scale than the selected tile size.
  In such cases, those tiles will be identified as tiles with no signal, over subtracting the Sky on the wings of bright sources, and creating artificial dark halos in the most extreme cases.
  An outlier identification algorithm has been added to \noisechisel{} to mitigate this problem since Gnuastro 0.8.
  It can be configured with the following options: \texttt{--outliersigma} and \texttt{--outliersclip}.

If we assume there are $N$ elements (useful tiles), the values are first sorted.
Searching for the outlier starts on element $N/2$ (integer division).
Let's take $v_i$ to be the $i$-th element of the sorted input (with no blank values) and $m$ and $\sigma$ as the $\sigma$-clipped median and standard deviation from the distances of the previous $N/2-1$ elements (not including $v_i$).
If the value given to \texttt{--outliersigma} is displayed with $s$, the $i$-th element is considered as an outlier when the condition below is true.

$${(v_i-v_{i-1})-m\over \sigma}>s$$

Since $i$ begins from the median, the outlier has to be larger than the median.
You can use the check images (for example \texttt{--checkqthresh}, \texttt{--checkdetsky} and \texttt{--checksky} options in \noisechisel{}) to inspect the steps and see which tiles have been discarded as outliers prior to interpolation.

\item
  \texttt{--blankasforeground}: allows blank pixels to be treated as foreground in \noisechisel's binary operations:
  the initial erosion (\texttt{--erode}) and opening (\texttt{--open}) as well as the filling holes and opening step for defining pseudo-detections (\texttt{--dthresh}).
  Until Gnuastro 0.8, \noisechisel{} treated blank pixels as foreground by default.
  But this could create false positives near blank/masked regions, blank pixels are now considered to be in the background by default.
  This option will re-create the old behavior.

\item
  \texttt{--skyfracnoblank}: To reduce the bias caused by undetected wings of galaxies and stars in the Sky measurements, \noisechisel{} only uses tiles that have a sufficiently large fraction of undetected pixels.
  Until Gnuastro 0.8, the reference for this fraction was the whole tile size.
  With this option, it is now possible to ask for ignoring blank pixels when calculating the fraction.
  This is useful when blank/masked pixels are distributed across the image.

\item
  \texttt{--dopening}: Number of openings after applying \texttt{--dthresh}, introduced in Gnuastro 0.9.

\item
  \texttt{--dopeningngb}: Type of connectivity to define neighbors (4-, or 8-connected) in \texttt{--dopening}, introduced in Gnuastro 0.9.

\item
\texttt{--holengb}: Type of connectivity to define neighbors (4-, or 8-connected) to define (and thus, fill) holes after applying \texttt{--dthresh} to find pseudo-detections. Introduced in Gnuastro 0.8.

\item
  \texttt{--pseudoconcomp}: Type of connectivity to define neighbors (4-, or 8-connected) to find individual pseudo-detections.
  For example if connectivity is defined by 8-connected neighbors, pseudo-detections that are touching on the corner will be identified as one.
  This was introduced in Gnuastro 0.9.

\item
\texttt{--snthresh}: Manually set the S/N of true pseudo-detections and thus avoid the need to automatically identify this value. Introduced in Gnuastro 0.9.

\item
  \texttt{--detgrowquant}: is used to grow the final true detections until a given quantile into the noise since Gnuastro 0.5.
  It replaces the old \texttt{--dilate} option in the paper and older versions of Gnuastro.
  Dilation is a blind growth method which causes objects to be boxy or diamond shaped when too many layers are added.
  However, with the growth method that is defined now, we can follow the signal into the noise with any shape.

  It uses a modified the Watershed algorithm, very similar to the way that clumps are grown during segmentation (compare columns 2 and 3 in Figure 10 of the paper).
  The difference with the Watershed algorithm is that when a pixel isn't touching any previously labeled pixel, it is not given a new label, it is just sent to the end of the sorted queue to be checked again in the next round.
  The best quantile to grow the quantile depends on your dataset's correlated noise properties and how cleanly it was Sky subtracted.
  The new \texttt{--detgrowmaxholesize} can also be used to specify the maximum hole size to fill as part of this growth.

  This new growth process can be much more successful in detecting diffuse flux around true detections compared to dilation and give more realistic results.
  For example see Figures \ref{detgrowquant} and \ref{detection}.
  Figure \ref{detgrowquant} shows how decreasing this option (going deeper into the noise) effects the final detection map.
  It can be seen that beyond a certain point, it will just start to follow correlated noise, emanating from the initial detections.
  See the discussion in Section \ref{m51demo} on how to address this problem while its not too significant.

\begin{figure}[t]
  \ifdefined\makepdf%
    \tikzsetnextfilename{detgrowquant}%
    \input{tex/src/detgrowquant.tex}%
  \else
    \includegraphics[width=\linewidth]{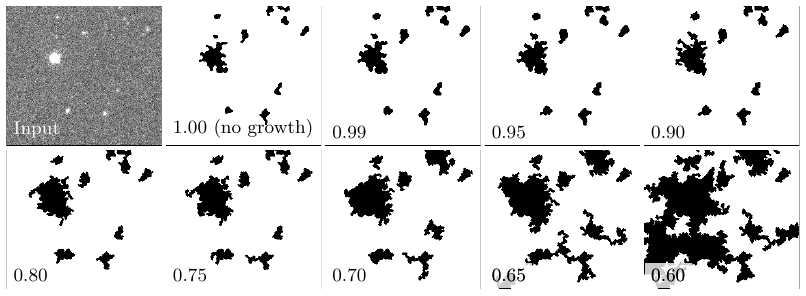}
  \fi

  \setlength\abovecaptionskip{-3mm}
  \caption{\label{detgrowquant}
    Growing true detections with various values to the \texttt{--detgrowquant} option in \noisechisel.
    The top-left image is the input image.
    Each subsequent binary image is the grown detection map after the growth has been completed to the given quantile.
    All other \noisechisel{} parameters are identical.
  }
\end{figure}

\item
  \texttt{--cleangrowndet}: After dilation, if the signal-to-noise ratio of a detection is less than the derived pseudo-detection \snsign{} limit, that detection will be discarded.
  In an ideal/clean noise, a true detection's \snsign{} should be larger than its constituent pseudo-detections because its area is larger and it also covers more signal.
  However, on false detections (especially at lower \texttt{--snquant} values), the increase in size can cause a decrease in S/N below that threshold.

  When this option is called, if the grown \snsign{} of a detected region is below the pseudo-detection \snsign{}, it is discarded.
  Because a true detection has flux in its vicinity and dilation will catch more of that flux and increase the S/N.
  This option will therefore improve purity and not effect the completeness (a true detection will not be discarded).

  However, in many real images bad processing creates artifacts that cannot be accurately removed by the Sky subtraction.
  In such cases, this option will decrease the completeness (will artificially discard true detections).
  So this feature is not default and should to be explicitly called when you know the noise is clean.
\end{itemize}

\section{\segment{} program for Segmentation}\label{updatedseg}

Prior to Gnuastro version 0.6 (released in June 2018), one program (\noisechisel) was in charge of detection \emph{and} segmentation.
To increase creativity and modularity, \noisechisel's segmentation features were spun-off into a separate program (\segment).
\segment's main algorithm and working strategy were initially defined and introduced in Section 3.2 of AI15.

\begin{itemize}

\item
  Since the spin-off from \noisechisel, the default kernel to smooth the input for convolution has a {\small FWHM} of 1.5 pixels (still a Gaussian).
  This is slightly less than \noisechisel's default kernel (which has a {\small FWHM} of 2 pixels).
  This enables the better detection of sharp/point-like clumps: as the kernel gets wider, the lower \snsign{} (but sharp/small) clumps will be washed away into the noise.
  Gnuastro's \textsf{MakeProfiles} can be used to build any custom kernel if this is too sharp/wide for a certain purpose.

  The ability to use a different convolution kernel for detection and segmentation is one example of how separating detection from segmentation into separate programs can increase productivity.
  The purpose of detection is separating diffuse and extended emission from noise, but in segmentation, sharp/localized peaks are desired.

\item
  The criteria to select true from false clumps is the peak significance.
  It is defined to be the difference between the clump's peak value ($C_c$) and the highest valued river pixel around that clump ($R_c$).
  Both are calculated on the convolved image (signified by the $c$ subscript).
  To avoid absolute values (differing from dataset to dataset), $C_c-R_c$ is then divided by the Sky standard deviation under the river pixel used ($\sigma_r$) as shown below (when \texttt{--minima} is given, the nominator becomes $R_c-C_c$):

$$C_c-R_c\over \sigma_r$$

  The input Sky standard deviation dataset (\texttt{--std}) is assumed to be for the unconvolved image.
  Therefore a constant factor (related to the convolution kernel) is necessary to convert this into an absolute peak significance\footnote{To get an estimate of the standard deviation correction factor between the input and convolved images, you can take the following steps:
    1) Mask (set to NaN) all detections on the convolved image with the \texttt{where} operator or Gnuastro's Arithmetic program.
    2) Calculate the standard deviation of the undetected (non-masked) pixels of the convolved image with the \texttt{--sky} option of Gnuastro's Statistics program (which also calculates the Sky standard deviation).
    Just make sure the tessellation settings of Statistics and \noisechisel{} are the same (you can check with the \texttt{-P} option).
    3) Divide the two standard deviation datasets to get the correction factor.}.
  However, as far as \segment{} is concerned, the absolute value of this correction factor is irrelevant: a purity level is found based on the ambient noise (undetected regions) to find the numerical threshold of this fraction.
  The user doesn't directly specify this value by default (unless \texttt{--clumpsnthresh} is given).

  A distribution's extremum (maximum or minimum) values, used in the new criteria, are strongly affected by scatter.
  On the other hand, the convolved image has much less scatter\footnote{For more on the effect of convolution on a distribution, see Section 3.1.1 of AI15}.
  Therefore $C_c-R_c$ is a more reliable (with less scatter) measure to identify signal than $C-R$ (on the unconvolved image).

  Initially, the total clump signal-to-noise ratio of each clump was used, see Section 3.2.1 of AI15.
  Therefore its completeness decreased dramatically when clumps were present on gradients for example see Figure \ref{segmentation}. In tests, this measure proved to be more successful in detecting clumps on gradients and on flatter regions simultaneously.

\item
With the new \texttt{--minima} option, it is now possible to detect inverse clumps (for example absorption features).
\end{itemize}

\section{Demonstration: M51 outskirts on a shallow image}\label{m51demo}

As a small demonstration of the detection and segmentation power of Gnuastro, a single exposure (exposure time of 54 seconds), SDSS \citep{sdss} image of M51 and NGC5195 in the \emph{r}-band is used.
It is a raw image directly downloaded when searching for NGC5195 in the SDSS ``simple field search'' online tool.

\begin{figure}[t]
  \ifdefined\makepdf%
    \tikzsetnextfilename{detection}%
    \input{tex/src/detection.tex}%
  \else
    \includegraphics[width=\linewidth]{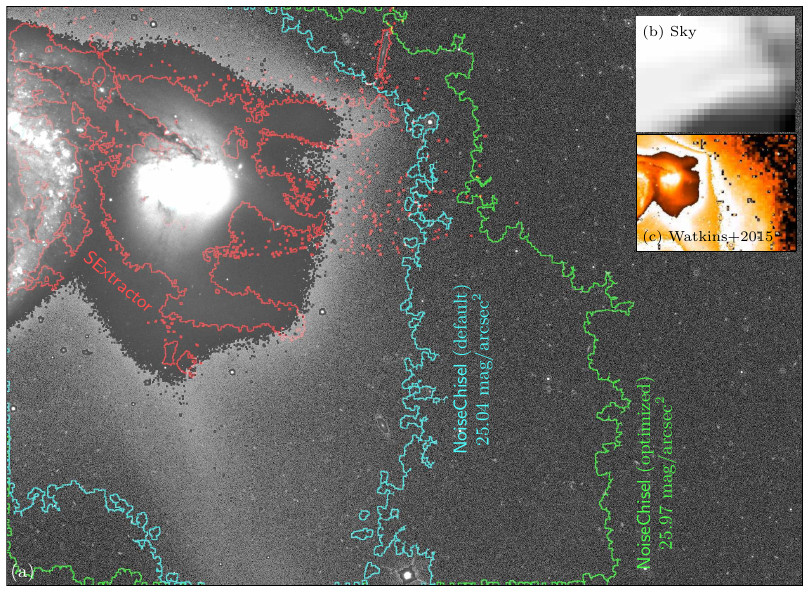}
  \fi

  \setlength\belowcaptionskip{1mm}
  \setlength\abovecaptionskip{-3mm}
  \caption{\label{detection}
    Detection of M51's outer wings on a single-exposure SDSS image.
    (a) Input image overlayed with the border of detected regions.
    The lower-\snsignsmaller{} regions of the image are shown in a logarithmic scale, the brighter parts are shown in linear scale to simultaneously show the familiar bright structure and outer low surface brightness wings.
    The red boundary shows \sextractor's detection (default parameters, except thresholds that are set to $\sedetectthresh\sigma$).
    The light-blue boundary shows the outer edge of the detection with \noisechisel's default configuration, and has an average \snsignsmaller{} of $\demosfdefaultsn$ (or $\demosfdefaultmagperarc$ mag/arcsec$^2$).
    The green boundary shows the result after optimizing the configuration for the noise of single-exposure {\scriptsize SDSS} images.
    The green boundary reaches an average \snsignsmaller{} of $\demosfoptimizedsn$ (or $\demosfoptimizedmagperarc$ mag/arcsec$^2$).
    For more on the customization of \sextractor{} and \noisechisel, see Section \ref{m51demo}.
    (b) Optimized \noisechisel{} Sky image, showing that there is still signal outside its detected regions.
    (c) Much deeper, processed image from \citet{watkins15} showing the outer signal at much higher \snsignsmaller, reproduced by permission of the {\scriptsize AAS}.
  }
\end{figure}

As a reference for readers who may be more familiar with \sextractor, it is run on the image first (version \sextractorversion).
The borders of \sextractor's detected pixels that overlap with M51 and NGC5195 are marked with red in Figure \ref{detection}.
The default value of all the \sextractor{} parameters were used (generated with the \texttt{-d} option) except for \texttt{DETECT\_THRESH=\sedetectthresh} and \texttt{ANALYSIS\_THRESH=\seanalysisthresh}.
The default thresholds of \texttt{1.5} (multiple of $\sigma$) were far too high and useless in this scenario.
Such a low threshold is rarely used because it creates many false positives (see Sections 5.2 and B.1.3 in AI15).

However, even with this decreased threshold, it can only detect the very bright parts of the signal.
This is primarily because of poor Sky subtraction in the vicinity of diffuse signal, see Figure 16 and Section A.6 of AI15.
All the small detections also had the same label as the large detections, similar to Figure \ref{uvudf}(2).
This is a big problem in a general catalog, where all these background galaxies will be missed in the final catalog.

\noisechisel{} (from Gnuastro \gnuastroversion) is also run on this image.
It is first run with its default parameters (no run-time options), producing the light-blue boundary in Figure \ref{detection} (\snsignsmaller=$\demosfdefaultsn$, or $\demosfdefaultmagperarc$ mag/arcsec$^2$).
However, the default parameters are only for a generic noise pattern.
Once \noisechisel{} is slightly tweaked for the single-exposure SDSS images (command below), it is able to detect the wings of M51 much more deeply as shown in the green border (\snsignsmaller=$\demosfoptimizedsn$, or $\demosfoptimizedmagperarc$ mag/arcsec$^2$, almost one magnitude deeper than the default):

{\small
  \texttt{astnoisechisel m51-r.fits -h0 --tilesize=\nctilesize{} --meanmedqdiff=\ncmeanmedqdiff{} \textbackslash}

  \texttt{{ }{ }{ }{ }{ }{ }{ }{ }{ }{ }{ }{ }{ }{ }{ }--interpnumngb=\ncinterpnumngb{} --snquant=\ncsnquant{} --detgrowquant=\ncdetgrowquant{} \textbackslash}

  \texttt{{ }{ }{ }{ }{ }{ }{ }{ }{ }{ }{ }{ }{ }{ }{ }--detgrowmaxholesize=\ncdetgrowmaxholesize{} --minskyfrac=\ncminskyfrac}
}

The displayed boundaries and surface brightness measurements of Figure \ref{detection} correspond to the \noisechisel{} output after being eroded two times.
The reason for this can be seen in Figure \ref{detgrowquant}.
As \texttt{--detgrowquant} is decreased, when there is no signal, \noisechisel{} starts following correlated noise that usually manifests as thin connections between smaller detections (like the three small detections that start touching beyond \texttt{--detgrowquant=0.65} in Figure \ref{detgrowquant}).
However, when there is signal (like the outer parts of the central galaxy), it is clear that \noisechisel{} is detecting very low \snsign{} signal and such thin branches aren't created.
Erosion removes the thin connections while not affecting the large contiguous diffuse detections.

Figure \ref{detection}(b) shows the Sky image (average of undetected pixels over a grid) for the optimized \noisechisel{} run.
Even though $\demosfoptimizedmagperarc$ mag/arcsec$^2$ seems very deep for a short-exposure image, the Sky image still resembles the shape of M51, showing that there is still fainter signal in this image.
In the tutorial, the readers are encouraged to optimize \noisechisel{} even further to obtain a better result.
Finally, Figure \ref{detection}(c) shows a rotated crop from the much deeper study of M51 by \citet{watkins15}, confirming the physical reality of this signal with the same morphology.

This is because \noisechisel{} has improved since Gnuastro \gnuastroversion, resulting in better and more optimal configurations.
Generally, this configuration is good for single-exposure SDSS images.
When dealing with a new type of dataset (images with different cameras, or reduced differently), for an optimal result, its always good to tweak it a little and optimize it for that particular noise-pattern (the logic is described in that tutorial).

\begin{figure}[t]
  \ifdefined\makepdf%
    \tikzsetnextfilename{segmentation}%
    \input{tex/src/segmentation.tex}%
  \else
    \includegraphics[width=\linewidth]{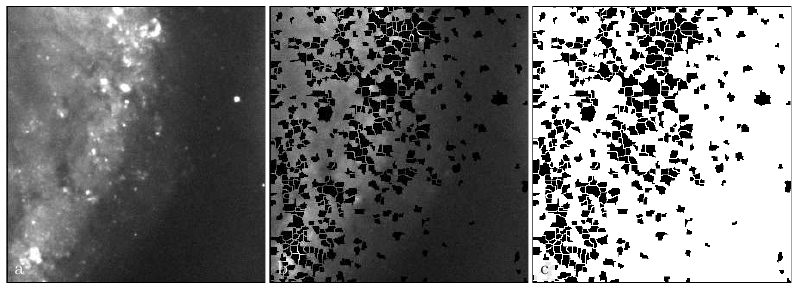}
  \fi

  \setlength\abovecaptionskip{-3mm}
  \setlength\belowcaptionskip{-1mm}
  \caption{\label{segmentation}
    Identification of HII regions over the M51 spiral arm in Figure \ref{detection} using \segment. (a) Input image, (b) all detected clumps masked, (c) the detected clumps.
  }
\end{figure}

Figure \ref{segmentation} shows the result of \segment{} (with its default configuration) on the brighter part of the M51 image.
It clearly shows the effectiveness of the new criteria to identify true peaks discussed in Section \ref{updatedseg}.
Each one of the HII regions in \segment's clumps can then be cataloged with \textsf{Make\-Catalog} to be studied individually, or masked to study the diffuse region.

It is important to highlight that parameters and optimizations above are for Gnuastro version 0.10 (released in August 2019).
\noisechisel{} has undergone several further improvements since then.
For a detailed list of updates, see the \texttt{NEWS} file\footnote{\url{https://git.savannah.gnu.org/cgit/gnuastro.git/plain/NEWS}}.
For a more up to date version of the M51 example shown here see the ``Detecting large extended targets'' tutorial of the Gnuastro manual\footnote{\scriptsize\url{https://www.gnu.org/s/gnuastro/manual/html_node/Detecting-large-extended-targets.html}} (which is always up to date).

\section{Reproducible data analysis}\label{repanalysis}

As shown in Figure \ref{detection}, when \noisechisel{} is optimized for a special noise pattern (in this case, any single-exposure SDSS image), it can operate significantly better.
But this isn't peculiar to \noisechisel{}, scientific software need to allow their users to modify every step of their analysis.
Some software authors try to decrease the number of tweak-able parameters by hard-coding the parameters within the code.
But this decreases the ability of the user (a scientist) to understand the details of the analysis and its effect on their scientific interpretation.
It is therefore critical for the integrity of the result that the authors report the exact details of all their software's configuration options, what order they were run, how they were installed and etc.
Simply reporting the name of a software is not enough.
But the format of a traditional paper like this also doesn't allow going into too much detail.

Generally, given the large volume of data already in surveys like {\small HSC} and {\small DES} (which will increase dramatically with next generation whole-sky surveys like J-PAS, Euclid or {\small LSST}), and the many configuration options (parameters to all the software used, software versions, and their operating environment), it is becoming harder to test new configurations, methods or software and thus understand the limits of the final choice in the low \snsign{} regime.
This is a major hurdle in using such studies for the low surface brightness universe (which is most sensitive).
Such tests can be facilitated/encouraged in a design where previous results can be exactly reproduced, on different machines, independent of the host operating system.

To address this major problem, AI15 also publicly released all the analysis scripts and configuration files that generated its results with the paper's \LaTeX{} source on arXiv.
Since that paper, that solution has grown into Maneage (\emph{Man}aging data lin\emph{eage}) (\url{https://maneage.org}), which provides a robust framework to start reproducible research projects \citep{maneage}.
Maneage installs fixed versions of all the necessary software from source and cites them where necessary (see Appendix \ref{appendix:software}).
Combined with all the scripts that produced every number or figure, this allows exact reproducibility of a paper by any random reader.

\begin{table}[t]
  \begin{center}
    \setlength\belowcaptionskip{1mm}
    \begin{tabular}{l c}
      \hline
      Analysis of Section \ref{m51demo}: & \href{https://archive.softwareheritage.org/swh:1:cnt:b92b52b759bab9c3debe8a9382644b9a84f15e7e;origin=https://gitlab.com/makhlaghi/iau-symposium-355;visit=swh:1:snp:1325f70b4ecf13ee193ff92f523f5f3547a12437;anchor=swh:1:rev:5db008710ee5bf3a0045f027202e6dfe9cd05c1d;path=/reproduce/analysis/make/m51.mk}{\texttt{swh:1:cnt:b92b52b759bab9c3debe8a9382644b9a84f15e7e}}\\

      \noisechisel{} options: & \href{https://archive.softwareheritage.org/swh:1:cnt:98345eb954dfee4c35e84e575e46b6b2dcdec50a;origin=https://gitlab.com/makhlaghi/iau-symposium-355;visit=swh:1:snp:1325f70b4ecf13ee193ff92f523f5f3547a12437;anchor=swh:1:rev:5db008710ee5bf3a0045f027202e6dfe9cd05c1d;path=/reproduce/analysis/config/noisechisel-options.conf}{\texttt{swh:1:cnt:98345eb954dfee4c35e84e575e46b6b2dcdec50a}}\\

      Input data identifiers: & \href{https://archive.softwareheritage.org/swh:1:cnt:25a79b02dc23ca88f48f52281bd1eb3cd1d3a6a9;origin=https://gitlab.com/makhlaghi/iau-symposium-355;visit=swh:1:snp:1325f70b4ecf13ee193ff92f523f5f3547a12437;anchor=swh:1:rev:5db008710ee5bf3a0045f027202e6dfe9cd05c1d;path=/reproduce/analysis/config/INPUTS.conf}{\texttt{swh:1:cnt:25a79b02dc23ca88f48f52281bd1eb3cd1d3a6a9}}\\
      \hline
    \end{tabular}
    \caption{\label{swhids}Software Heritage IDs (SWHIDs, click-able) for direct access to some parts of this paper's analysis that is exactly reproducible, using the Maneage framework, see text for more.
    If clicking is not available (for example in print), the SWHIDs can be resolved by prefixing them with `https://n2t.net/' (for example `https://n2t.net/swh:1:cnt:...')}
  \end{center}
\end{table}

This paper has been written using Maneage and can be exactly reproduced, verified or modified by retrieving version \texttt{\projectversion} of its history\footnote{\url{https://gitlab.com/makhlaghi/iau-symposium-355}}.
The repository is also archived on SoftwareHeritage \citep{softwareheritage}, therefore, as shown in Table \ref{swhids}, parts of the project are accessible via the click-able SWHIDs \citep{swhid}.
The reproducible source includes links to download the input images from their respective server, as well as their MD5 checksum to confirm that the downloaded file is the same that was used here (has not been updated on the server).
All necessary files (reproducible source, its Git history and software tarballs) are also archived on Zenodo\footnote{\url{\zenododoi}}.

\section{Acknowledgments}

This research was done using GNU Astronomy Utilities (Gnuastro, ascl.net/1801.009), and in the reproducible framework of Maneage \citep[\emph{Man}aging data lin\emph{eage},][latest Maneage commit \texttt{\maneageversion}, from \maneagedate]{maneage}.
The project was built on an {\machinearchitecture} machine with {\machinebyteorder} byte-order, see Appendix \ref{appendix:software} for the used software and their versions.
Work on Gnuastro and Maneage has been funded by the Japanese Ministry of Education, Culture, Sports, Science, and Technology (MEXT) scholarship and its Grant-in-Aid for Scientific Research (21244012, 24253003), the Spanish Ministry of Economy and Competitiveness (MINECO) under grant number AYA2016-76219-P, and the European Union (EU) ERC advanced grant 339659 (MUSICOS) and EU MSCA-ITN grant 721463 (SUNDIAL).
Maneage was also awarded a dedicated cascading grant (PI: M. Akhlaghi) from EU INFRASUPP grant 777388 (RDA Europe 4.0), under the RDA Europe Adoption Grants of 2019.

This research was done using data from SDSS.
Funding for the SDSS and SDSS-II has been provided by the Alfred P. Sloan Foundation, the Participating Institutions, the National Science Foundation, the U.S. Department of Energy, the National Aeronautics and Space Administration, the Japanese Monbukagakusho, the Max Planck Society, and the Higher Education Funding Council for England.
The SDSS Web Site is http://www.sdss.org.

\printbibliography

\newpage
\appendix

\section{Software acknowledgment}
\label{appendix:software}
 
This research was done with the following free software programs and libraries: Bzip2 1.0.8, CFITSIO 3.48, CMake 3.21.4, cURL 7.79.1, Dash 0.5.11.5, Discoteq flock 0.4.0, Expat 2.4.1, FFTW 3.3.10 \citep{fftw}, File 5.41, Fontconfig 2.13.94, FreeType 2.11.0, Git 2.36.0, GNU Astronomy Utilities 0.10 \citep{gnuastro}, GNU Autoconf 2.71, GNU Automake 1.16.5, GNU AWK 5.1.0, GNU Bash 5.1.8, GNU Binutils 2.37, GNU Compiler Collection (GCC) 11.2.0, GNU Coreutils 9.1, GNU Diffutils 3.8, GNU Findutils 4.8.0, GNU gettext 0.21, GNU gperf 3.1, GNU Grep 3.7, GNU Gzip 1.11, GNU Integer Set Library 0.18, GNU libiconv 1.16, GNU Libtool 2.4.6, GNU libunistring 1.0, GNU M4 1.4.19, GNU Make 4.3, GNU Multiple Precision Arithmetic Library 6.2.1, GNU Multiple Precision Complex library, GNU Multiple Precision Floating-Point Reliably 4.1.0, GNU Nano 6.0, GNU NCURSES 6.3, GNU Readline 8.1.1, GNU Scientific Library 2.7, GNU Sed 4.8, GNU Tar 1.34, GNU Texinfo 6.8, GNU Wget 1.21.2, GNU Which 2.21, GPL Ghostscript 9.55.0, Less 590, Libffi 3.4.2, Libgit2 1.3.0, libICE 1.0.10, Libidn 1.38, Libjpeg 9d, Libpaper 1.1.28, Libpng 1.6.37, libpthread-stubs (Xorg) 0.4, libSM 1.2.3, Libtiff 4.3.0, libXau (Xorg) 1.0.9, libxcb (Xorg) 1.14, libXdmcp (Xorg) 1.1.3, libXext 1.3.4, Libxml2 2.9.12, libXt 1.2.1, Lzip 1.22, OpenBLAS 0.3.18, OpenSSL 3.0.0, PatchELF 0.13, Perl 5.34.0, pkg-config 0.29.2, podlators 4.14, Python 3.10.0, SExtractor 2.25.0 \citep{sextractor}, Unzip 6.0, util-Linux 2.37.2, util-macros (Xorg) 1.19.3, WCSLIB 7.7, X11 library 1.7.2, XCB-proto (Xorg) 1.14.1, xorgproto 2021.5, xtrans (Xorg) 1.4.0, XZ Utils 5.2.5, Zip 3.0 and Zlib 1.2.11. 
The \LaTeX{} source of the paper was compiled to make the PDF using the following packages: biber 2.17, biblatex 3.17, bitset 1.3, caption 62757 (revision), courier 61719 (revision), csquotes 5.2l, datetime 2.60, ec 1.0, etoolbox 2.5k, fancyhdr 4.0.1, fmtcount 3.07, fontaxes 1.0e, footmisc 6.0d, fp 2.1d, kastrup 15878 (revision), letltxmacro 1.6, logreq 1.0, mweights 53520 (revision), newtx 1.71, pdfescape 1.15, pdftexcmds 0.33, pgf 3.1.9a, pgfplots 1.18.1, preprint 2011, setspace 6.7a, tex 3.141592653, texgyre 2.501, times 61719 (revision), titlesec 2.14, trimspaces 1.1, txfonts 15878 (revision), ulem 53365 (revision), xcolor 2.13, xkeyval 2.8 and xstring 1.84. 
We are very grateful to all their creators for freely  providing this necessary infrastructure. This research  (and many other projects) would not be possible without  them.

\section{Corrections after first submission}
\label{appendix:corrections}
In the submission to the Proceedings of IAU Symposium No. 355, and first arXiv upload (Commit \texttt{\firstsubmitcommit}, from \firstsubmitdate), the SDSS Asinh Magnitude zero points were mistakenly used ($\sdsszpasinhr$ in the r filter\footnote{\url{https://www.sdss.org/dr14/algorithms/magnitudes}}).
However, the downloaded SDSS image was in units of Nanomaggies (from the FITS keyword \texttt{BUNIT}), which has a fixed zero point of $\sdsszpnanomaggie$ (the zero-point was not written in the SDSS FITS file).
Therefore the reported outer surface brightness levels of M51 were overestimated by $\sdsszpdiff$ mag/arcsec$^2$.
This issue has been corrected in the current version and reported to the proceedings editors.

The project was also merged with the \texttt{maneage} branch to update its low-level infrastructure and add automatic verification (which wasn't available on the first submission of this paper).
Some low-level dependency versions have changed in this update, but the versions of Gnuastro and \sextractor{} (the main analysis software) haven't changed and except for the zero point-affected results, no other result has been affected.

\end{document}

%
%
%